\documentclass[manuscript,screen]{acmart}

\usepackage[most]{tcolorbox}
\usepackage{booktabs}

\newcommand{\tosemtablewidth}{0.96\textwidth}

\acmJournal{TOSEM}
\setcopyright{none}
\acmDOI{}
\renewcommand\footnotetextcopyrightpermission[1]{}

\providecommand{\doi}[1]{\url{https://doi.org/#1}}
\newtcolorbox{observationbox}{
  colback=black!6,
  colframe=black,
  boxrule=0.5pt,
  arc=3pt,
  left=6pt,
  right=6pt,
  top=5pt,
  bottom=5pt,
  before skip=0.6em,
  after skip=0.6em,
}
\newtcolorbox{answerbox}{
  enhanced,
  colback=black!6,
  boxrule=0pt,
  frame hidden,
  sharp corners,
  borderline west={2.5pt}{0pt}{black!55},
  left=8pt,
  right=6pt,
  top=6pt,
  bottom=6pt,
  before skip=0.8em,
  after skip=0.8em,
}
\newtcolorbox{issuebox}{
  colback=black!4,
  colframe=black!55,
  boxrule=0.4pt,
  arc=2pt,
  left=5pt,
  right=5pt,
  top=4pt,
  bottom=4pt,
  fontupper=\footnotesize\ttfamily,
}

\begin{document}

\title{The Web4 Agent Economy: A Large-Scale Empirical Study of the Landscape, Challenges, and Opportunities}

\author{Yuhan Jin}
\email{valkqs@gmail.com}
\affiliation{%
  \department{The State Key Laboratory of Blockchain and Data Security}
  \institution{Zhejiang University}
  \city{Hangzhou}
  \country{China}}

\author{Shuohan Wu}
\affiliation{%
  \department{The State Key Laboratory of Blockchain and Data Security}
  \institution{Zhejiang University}
  \city{Hangzhou}
  \country{China}}
\affiliation{%
  \institution{Hangzhou High-Tech Zone (Binjiang) Institute of Blockchain and Data Security}
  \city{Hangzhou}
  \country{China}}

\author{Chong Chen}
\affiliation{%
  \department{The State Key Laboratory of Blockchain and Data Security}
  \institution{Zhejiang University}
  \city{Hangzhou}
  \country{China}}

\author{Lingfeng Bao}
\affiliation{%
  \department{The State Key Laboratory of Blockchain and Data Security}
  \institution{Zhejiang University}
  \city{Hangzhou}
  \country{China}}
\affiliation{%
  \institution{Hangzhou High-Tech Zone (Binjiang) Institute of Blockchain and Data Security}
  \city{Hangzhou}
  \country{China}}

\author{Xiaohu Yang}
\affiliation{%
  \department{The State Key Laboratory of Blockchain and Data Security}
  \institution{Zhejiang University}
  \city{Hangzhou}
  \country{China}}
\affiliation{%
  \institution{Hangzhou High-Tech Zone (Binjiang) Institute of Blockchain and Data Security}
  \city{Hangzhou}
  \country{China}}

\author{Jiachi Chen}
\authornote{Corresponding author.}
\email{jiachi.chen@zju.edu.cn}
\affiliation{%
  \department{The State Key Laboratory of Blockchain and Data Security}
  \institution{Zhejiang University}
  \city{Hangzhou}
  \country{China}}
\affiliation{%
  \institution{Hangzhou High-Tech Zone (Binjiang) Institute of Blockchain and Data Security}
  \city{Hangzhou}
  \country{China}}

\renewcommand{\shortauthors}{Jin et al.}

\begin{abstract}

With the rapid advancement of AI agents, the Internet is evolving from Web3 toward Web4, where autonomous agents serve as independent economic actors. 
These agents can control crypto wallets, execute on-chain transactions, and pay for external API services.
This transition calls for a new infrastructure stack that supports key agent operations, including agent-to-tool interaction, agent-to-agent payments, and verifiable agent identity. These capabilities are enabled by emerging protocols such as the Model Context Protocol (MCP), x402, and EIP-8004, respectively.
Despite growing industrial interest in these protocols, the real-world Web4 agent ecosystem remains largely underexplored.
To bridge this gap, we conduct the \textit{first} large-scale empirical study of the Web4 ecosystem. 
Specifically, our study targets three research questions: (1) how the Web4 agent economy has emerged in practice; (2) what practical challenges arise in Web4 agent development; and (3) how current projects address these challenges.
To answer these questions, we analyze 386,665 EIP-8004 identities, 11.64 million x402 settlements, the transaction histories of 605 sampled MCP-linked wallets, and 377 classified GitHub issues from 75 repositories.
By analyzing these dimensions, we characterize the emerging Web4 ecosystem, identify key security and reliability gaps, and provide practical directions for building safer and more trustworthy agent systems.

\end{abstract}



\maketitle

\section{Introduction}
\label{sec:01_Introduction}
The past decade has witnessed the development of Web3, where blockchain technology enabled decentralized ownership of user data and assets~\cite{huang2024overview}. 
With the rise of LLM-powered agents, the Internet is now evolving toward Web4. Web4 is the next-generation internet that integrates AI and blockchain to enable autonomous agents to reason, make decisions, and participate in on-chain economies as independent economic actors~\cite{kilbourn2023intent,gurpinar2025web4}.

To make such agent-driven economies possible, autonomous agents must be able to turn decisions into executable and accountable economic actions.
For example, a Web4 trading agent can invoke external tools to obtain real-time market data, execute trades through decentralized exchanges, pay data providers for the resources it consumes, and expose verifiable identity and reputation for other agents or users to assess before engaging~\cite{liu2025aiagents}. Supporting this workflow requires new infrastructure for standardized agent-to-tool interaction, programmable agent-to-agent payments, and verifiable agent identity. 

Emerging protocols have begun to provide these capabilities. The \textit{Model Context Protocol (MCP)} standardizes agent-to-tool interaction by allowing AI agents to discover and invoke external tools, APIs, and on-chain smart contracts through a unified interface~\cite{anthropic2024mcp}. The \textit{x402 protocol} enables programmable machine-to-machine payments, allowing agents to pay for fine-grained resources such as API access~\cite{coinbase2025x402}. 
\textit{EIP-8004} provides a decentralized identity mechanism for autonomous agents and links an agent's capabilities, reputation, and validation records with a verifiable on-chain identifier~\cite{eip8004trustless}.
Together, these protocols constitute the core infrastructure of the Web4 ecosystem, enabling agents to access tools, exchange value, and present verifiable identities in open networks.

However, despite growing interest in Web4, the real-world Web4 ecosystem remains largely unexplored. Existing studies~\cite{nath2015comes,zhou2023review,kobayashi2026web4,palatov2025web4} mainly focus on its conceptual definitions, architectural designs, and prospective application scenarios. It remains unclear whether the emerging Web4 ecosystem is technically mature enough to support secure and trustworthy agent-driven economies. This motivates a comprehensive empirical analysis that can help both academia and industry understand the current ecosystem landscape and identify its major challenges and risks.

In this paper, we conduct the \textbf{\textit{first}} large-scale empirical study of the Web4 ecosystem. We examine its real-world infrastructure and practical challenges by answering the following three research questions.

\noindent \textit{\textbf{RQ1: How has the Web4 agent economy emerged in practice?}}

To characterize where Web4 agents are deployed, how they participate in machine-to-machine payments, and what types of on-chain capabilities they currently expose, we analyze 386,665 EIP-8004 identities and 20,922 identities that publish MCP services across 26 active blockchain mainnets, together with 11.64 million x402 settlements on Base~\cite{coinbase2025x402} over 30 days and 605 sampled MCP-linked wallets. The results reveal a considerable but uneven ecosystem. First, the networks leading in identity registration, reputation activity, and MCP publication are different, showing that growth across these layers is not synchronized. Second, x402 exhibits high-throughput, low-value settlements across a broad seller base relative to buyers, indicating a market centered on repeated, fine-grained service exchange. Third, MCP adoption shows a divide between participation breadth and activity intensity: broader service publication does not necessarily translate into broader wallet participation, while activity intensity can remain high even when participation is narrow.

\noindent \textit{\textbf{RQ2: What practical challenges arise in Web4 agent development?}}

Web4 agents can control external tools, wallets, and digital assets; flaws introduced during development can therefore disrupt services or cause direct economic losses. We thus systematically examine developer-reported issues in current Web4 projects. To this end, we collect 2,275 issues from 117 Web4-related repositories and retain 882 after relevance screening and deduplication. We first define seven issue categories based on the Web4 development lifecycle and established classifications of software defects and engineering concerns. We then use LLMs to independently classify each retained issue into these categories and extract supporting evidence. This process identifies 377 issues across the seven categories and assigns 394 labels because some issues span multiple categories. We find that Web4 development is hindered primarily by missing security controls and inadequate documentation, build support, and diagnostic information. We further identify Web4-specific failures at the boundaries of delegated authorization, on-chain execution, and payment settlement.

\noindent \textit{\textbf{RQ3: How are current Web4 projects addressing reported  challenges?}}

We analyze how current Web4 projects respond to reported development issues and which response strategies they use, which can help researchers, developers, and maintainers identify effective response practices, prioritize unresolved issues, and improve the reliability of Web4 systems. For 376 of the 377 issues classified in RQ2, we retrieve issue discussions and linked implementation evidence, including pull requests, reviews, commits, and changed files. We then use LLMs to determine each issue's response status and classify its response methods. The results show that just over half of the issues (205, 54.5\%) received a substantive response. We further identify distinct response patterns for Web4-specific issues: projects address authorization failures by isolating signing authority and adding permission or confirmation gates; chain-runtime failures by filtering stale state, correcting transaction inputs, and adding retry or backoff; and payment failures by blocking service delivery until settlement succeeds.

In this paper, we make the following contributions:

\begin{itemize}

\item We present the \textit{first} large-scale empirical study of the Web4 agent economy using 386,665 EIP-8004 identities, 527,270 Feedback records, 11.64 million x402 settlements, and transaction histories for 605 sampled MCP-linked wallets. The results show that identity registration, service publication, machine payments, and on-chain execution are operational, but concentrate on different networks.

\item We construct a seven-category taxonomy of practical Web4 development challenges from 377 relevant GitHub issues. Missing security controls are reported most often, followed by documentation, dependency/build, and monitoring and diagnosis problems; Web4-specific failures occur at authorization, on-chain execution, and payment-settlement boundaries.

\item We examine project responses by analyzing discussions, pull requests, reviews, commits, and changed files for 376 issues. We identify 205 substantive responses, while 82 issues receive only administrative action and 78 have no observable response; Web4-specific responses focus on signing and permission checks, chain-facing execution, and payment-before-delivery enforcement.

\end{itemize}

\section{Background}
\label{sec:02_Background}

\subsection{Web4}
The transition from Web3 to Web4 changes the role of AI from an off-chain analytical tool to an autonomous participant in blockchain systems~\cite{gurpinar2025web4}. Web4 extends Web3 by introducing AI agents that can interpret user goals and act across blockchain applications~\cite{gurpinar2025web4,kilbourn2023intent}. Web4 adopts an intent-centric model in which users specify goals and agents determine how to execute them~\cite{kilbourn2023intent}. Large Language Models (LLMs) provide the reasoning and action-planning capabilities required for this role~\cite{yao2022react,wang2024survey}. Agents can then translate abstract goals into tool calls and on-chain operations, including machine-to-machine payments~\cite{coinbase2025x402}.

To achieve this autonomy, the infrastructure must integrate non-deterministic LLM reasoning~\cite{yao2022react,wang2024survey} with deterministic blockchain environments~\cite{wood2014ethereum}. This integration is primarily supported by an emerging tri-layer architecture consisting of the economic protocol (x402)~\cite{coinbase2025x402}, the execution standard (MCP)~\cite{anthropic2024mcp}, and the identity/trust registry (EIP-8004)~\cite{eip8004trustless}. We next introduce these three protocols.

\subsection{The x402 Protocol}
To be truly independent, AI agents must be able to earn revenue and pay for external resources, such as API calls or data storage~\cite{coinbase2025x402,coinbase2025agentkit}. Traditional Web2 payment rails are poorly suited to high-frequency Machine-to-Machine (M2M) micropayments: they settle slowly~\cite{stripe2026settlement} and impose fixed per-transaction fees that dominate small amounts~\cite{visa2026interchange,stripe2026pricing}. For example, Visa's U.S.\ interchange schedule lists a flat \$0.30 fee for certain supermarket debit transactions~\cite{visa2026interchange}; on a \$0.31 payment, that fixed fee alone nearly cancels the transfer. Merchant-facing processors show the same structure (e.g., Stripe's standard U.S.\ online card rate of 2.9\% + \$0.30)~\cite{stripe2026pricing}, while card settlement commonly takes one to three business days~\cite{stripe2026settlement}. Agents that need to pay cents (or less) per API call therefore cannot rely on card networks.

The HTTP 402 ``Payment Required'' status code~\cite{rfc7231,lightning2020l402} inspired x402, which provides a programmable payment path for Web3/Web4 agents~\cite{coinbase2025x402}. Using stablecoins (e.g., USDC)~\cite{circle2024usdc,osl2026monthly,talos2026state}, x402 allows agents to perform automated payments in code-driven workflows. While early IoT research discussed M2M economies in theory~\cite{christidis2016blockchains}, large-scale empirical evidence in Web4 contexts remains limited, which motivates our ecosystem-level measurement and obstacle analysis.

\subsection{Model Context Protocol (MCP)}
When LLMs act as independent agents~\cite{yao2022react,wang2024survey}, they need a standard way to invoke tools, APIs, and smart contracts. The Model Context Protocol (MCP) has become a widely used standard for this purpose~\cite{anthropic2024mcp}. MCP provides a unified interface that enables models to discover and use external capabilities without bespoke integration for each tool~\cite{anthropic2024mcp,hou2025mcp}.

In Web4 settings, MCP functions as an execution interface between AI reasoning and blockchain operations~\cite{anthropic2024mcp,coinbase2025agentkit}. Existing studies have begun to examine MCP usage and related risks~\cite{hasan2025mcp,hou2025mcp,greshake2023more,owasp2023top10llm}. These concerns motivate our empirical analysis of practical issues and response patterns in real projects.

\subsection{EIP-8004 and Agent Identity}
In open Web4 networks where anyone can launch AI agents, identity authenticity and accountability become central concerns~\cite{gurpinar2025web4,palatov2025web4}. Generic decentralized identifiers (DIDs) were originally designed for broader identity contexts~\cite{sporny2022did}, while EIP-8004 targets autonomous agent scenarios more directly~\cite{eip8004trustless,8004scan2026explorer}.

EIP-8004 links agent addresses with capability and trust-related records on-chain, and provides protocol-level structures for identity-oriented coordination~\cite{eip8004trustless,8004scan2026explorer}. In this paper, we treat this identity layer as an empirical part of ecosystem infrastructure and analyze it jointly with execution and payment layers to understand current practice~\cite{anthropic2024mcp,coinbase2025x402,eip8004trustless}.

\section{RQ1: Web4 Agent Economy in Practice}
\label{sec:03_RQ1_MacroscopicTopology}
In this section, we measure three observable parts of the Web4 agent economy, \textit{i.e.,} EIP-8004 registered agent identities, x402 payments, and transactions linked to MCP services. These data show the size of the registered agent population, the current scale of machine payments, and the types of on-chain operations performed by MCP.

\subsection{Methodology}

\begin{figure}[t]
\centering
\includegraphics[page=1,width=\textwidth]{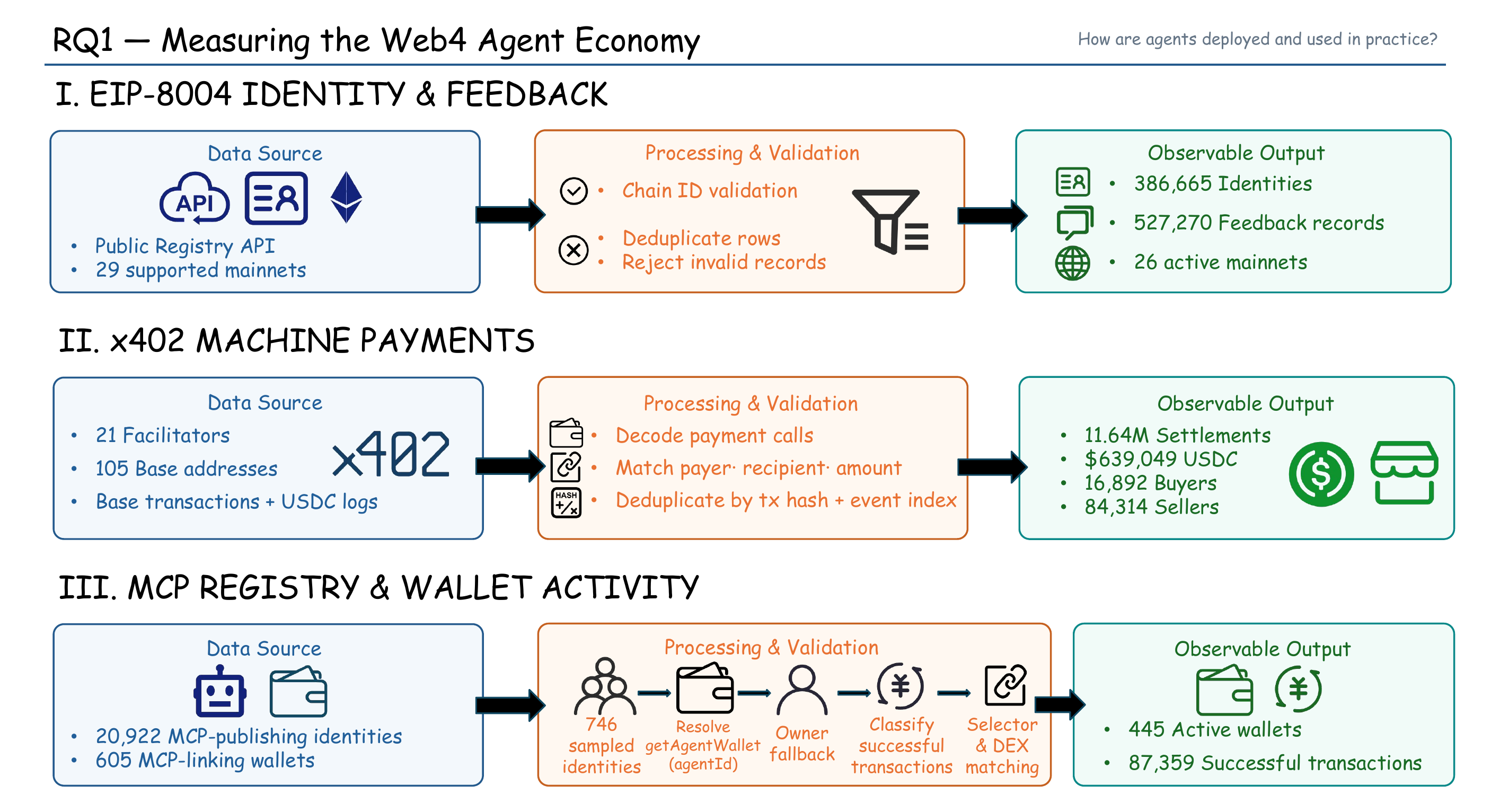}
\caption{Overview of data collection, validation, and measurement for EIP-8004 identities, x402 payments, and MCP-linked wallet activity.}
\Description{Three parallel measurement pipelines. The first validates registry data and reports 386,665 identities and 527,270 Feedback records across 26 active mainnets. The second decodes and deduplicates Base USDC payment records and reports 11.64 million x402 settlements. The third resolves wallets linked to MCP-publishing identities, classifies successful transactions, and reports 445 active wallets and 87,359 transactions.}
\label{fig:rq1-methodology-overview}
\end{figure}

\noindent \textbf{Overview.} We collected the EIP-8004 registry, x402 transaction, and MCP publication data on August 3, 2026. The three measurements cover registered identities and Feedback, completed machine payments, and on-chain activity linked to MCP publishers. Figure~\ref{fig:rq1-methodology-overview} summarizes the data sources, validation steps, and metrics used for these measurements. Their network scopes and measurement rules are described below.

\subsubsection{EIP-8004 Identity}

We collect and compare identity registrations and Feedback records to measure registry use across mainnets.

\noindent \textbf{Data collection and validation.} We examine the scale and distribution of EIP-8004 registrations and Feedback records across all supported mainnets returned by the public registry interface. We developed an automated collector which will first retrieve the complete list of supported networks and their chain IDs. At a common cutoff, it then retrieves one aggregate statistics object containing a record for every network and extracts the same registration and Feedback fields from each record.

The collector uses the chain ID as the unique network key, removes duplicate rows, and rejects records with missing identifiers or invalid counts. It then sums the per-network counts, checks the number of supported and active networks, and ranks networks by registered identities. This process produces 29 supported mainnets, including 26 with at least one registered identity. For each mainnet, the collector records the chain ID, network name, total number of registered identities, and total number of Feedback records. Under EIP-8004, a Feedback record is an on-chain reputation entry associated with a registered identity. We use Feedback as evidence of interaction with the EIP-8004 reputation mechanism. Test networks are excluded, while mainnets with zero registrations are retained to preserve the full deployment scope.

\noindent \textbf{Metrics and network comparison.} We calculate the total number of identities, the number of networks with at least one identity, and each network's share of both totals. We also divide Feedback records by registered identities on each active network. This measure shows whether a large registered population is accompanied by comparable use of the reputation mechanism. After comparing all supported mainnets, we examine BNB Smart Chain, Base, and Ethereum~\cite{bnbchain2026bsc,base2026overview,wood2014ethereum} in detail because they rank first, second, and third by registered identities.

\subsubsection{x402 Settlement Measurement}

We identify completed x402 payments on Base to measure their scale, value, and participants.

\noindent \textbf{Measurement scope.} The x402 analysis focuses on USDC settlements on Base. We select USDC because it is the default payment asset in the x402 reference workflow and provides standardized on-chain records for consistent identification and valuation. We use the Base addresses published by known facilitators to link these transfers to x402. At the collection cutoff, the public index lists 21 facilitators and 105 Base addresses~\cite{x402scan2026index,coinbase2025x402,circle2024usdc}.

\noindent \textbf{Settlement identification.} In a direct settlement, the facilitator submits the payer's off-chain-signed USDC authorization through EIP-3009's \texttt{transferWithAuthorization}; this function is therefore the observable on-chain entry point of the standard x402 payment flow. We identify a settlement in three steps. First, we collect Base transactions sent by the published facilitator addresses or handled by their settlement contracts. Second, we decode the payment fields in direct \texttt{transferWithAuthorization} calls or in supported contract calls. Third, we require a Base USDC \texttt{Transfer} event with the same payer, recipient, and amount. We remove duplicates using the transaction hash and event index. Ordinary USDC transfers that do not pass these checks are excluded.

\subsubsection{MCP Publication and On-chain Capability Measurement}

We link MCP publishers to their wallets and analyze wallet activity to measure their on-chain use.

\noindent \textbf{Service identification.} After measuring identity registration and x402 settlement activity, we examine how many identities publish MCP services, whether these identities are linked to active wallets, and what operations those wallets send. We classify an EIP-8004 identity as an MCP publisher if its registry metadata lists an MCP service. The \emph{proportion of identities publishing MCP services} is the number of these identities divided by all registered identities on the same network. This measure describes how often an on-chain identity makes an MCP service available. We examine Celo~\cite{celo2026overview}, Base, and BNB Smart Chain in detail because they have the three largest populations of identities that publish MCP services.

\noindent \textbf{Identity sampling and wallet linking.} Publishing an MCP service does not reveal whether the publisher is active on-chain. We therefore analyze transactions from its EIP-8004-linked agent wallet as observable evidence of economic and contract activity. To study the wallets linked to these identities, we sample MCP records randomly and separately from each network. After duplicate records are removed, the sample contains 746 identities.

An EIP-8004 identity and the wallet that acts for it are not necessarily the same. An identity may be created and held by a registry owner, while a separate agent wallet performs its on-chain operations. For each sampled identity, we therefore call \texttt{getAgentWallet(agentId)} on the IdentityRegistry. We use the returned non-zero address as the linked wallet. If the call returns zero, we use the recorded owner address as a fallback and do not treat it as evidence of a separate agent wallet. After duplicate addresses are removed, the 746 identities correspond to 605 distinct wallets.

\noindent \textbf{Wallet activity measurement.} We query Dune~\cite{dune2026analytics} for successful transactions sent by these wallets from January 1 to the collection time on August 1, 2026. Filtering by sender ensures that the sampled wallet initiated the operation, while retaining only successful transactions excludes reverted attempts. A wallet is active if it sends at least one such transaction. We calculate the share of active wallets and the number of transactions per active wallet. These measures separate broad participation from activity concentrated in a small number of wallets.

\noindent \textbf{Operation classification.} Finally, to characterize what MCP-linked wallets do on-chain, we classify each transaction using the beginning of its call data. We map known selectors to ERC-20 transfer, token approval, transfer-from call, lending call, or DEX call, while transactions without call data are treated as direct transfers of a network's native token. Some trades use routers or aggregators whose top-level selectors do not reveal the underlying operation. For transactions that remain unclassified after selector-based mapping, we first match their hashes against Dune's \texttt{dex.trades} records. We then reclassify the matched transactions as DEX transactions. We use distinct transaction hashes before matching, so a trade executed through multiple steps is counted once. The resulting categories are mutually exclusive, and we report their shares for each network.

\subsection{Results}

\subsubsection{EIP-8004 Registry Deployment and Feedback Activity}

\begin{table}[t]
\caption{Distribution of EIP-8004 identities and Feedback records across selected mainnets and the remaining active networks}
\label{tab:eip8004_identity_feedback}
\centering
\begingroup
\small
\setlength{\tabcolsep}{5.5pt}
\renewcommand{\arraystretch}{1.12}
\begin{tabular*}{\tosemtablewidth}{@{\extracolsep{\fill}}lrrrrr@{}}
\toprule
& \multicolumn{2}{c}{\bfseries Registry} & \multicolumn{3}{c}{\bfseries Feedback activity} \\
\cmidrule(lr){2-3}\cmidrule(l){4-6}
{\bfseries Mainnet} & {\bfseries Identities} & {\bfseries Share} & {\bfseries Records} & {\bfseries Share} & {\bfseries Per identity} \\
\midrule
BNB Smart Chain & 238,206 & 61.61\% & 11,705 & 2.22\% & 0.049 \\
Base & 40,591 & 10.50\% & 418,959 & 79.46\% & 10.32 \\
Ethereum & 29,172 & 7.54\% & 3,209 & 0.61\% & 0.11 \\
Celo & 9,701 & 2.51\% & 27,091 & 5.14\% & 2.79 \\
Other 22 active mainnets & 68,995 & 17.84\% & 66,306 & 12.58\% & 0.96 \\
\midrule
{\bfseries Total} & {\bfseries 386,665} & {\bfseries 100.00\%} & {\bfseries 527,270} & {\bfseries 100.00\%} & {\bfseries 1.36} \\
\bottomrule
\end{tabular*}
\endgroup
\end{table}

Table~\ref{tab:eip8004_identity_feedback} summarizes the distribution of EIP-8004 identities and Feedback records across active mainnets. EIP-8004 identity registration and reputation activity have developed at substantially different rates across networks. Across the 26 active mainnets, EIP-8004 has accumulated 386,665 registered identities and 527,270 Feedback records. However, these two forms of activity follow markedly different distributions: BNB Smart Chain hosts 61.61\% of all registered identities but accounts for only 2.22\% of Feedback records, whereas Base hosts only 10.50\% of the identities but accounts for 79.46\% of Feedback. This contrast shows that large-scale identity registration does not necessarily translate into active use of the reputation mechanism. Instead, identity deployment and reputation activity are concentrated on different networks, revealing a fragmented pattern of EIP-8004 adoption.

\textit{First}, identity registration is highly concentrated on a small number of networks. The data contain 386,665 registered identities across 29 supported mainnets, of which 26 contain at least one identity. BNB Smart Chain alone contains 238,206 identities, or 61.61\% of all registrations. Base ranks second with 40,591 identities (10.50\%), followed by Ethereum with 29,172 (7.54\%). These three networks contain 79.65\% of all registered identities, while the other 23 active networks share the remaining 20.35\%; the median active network contains about 1,040 identities.

\textit{Second}, the distribution of reputation activity differs markedly from that of registered identities. The data contain 527,270 Feedback records, of which Base holds 418,959 (79.46\%) despite accounting for only 40,591 identities (10.50\%). In contrast, BNB Smart Chain contains 238,206 identities (61.61\%) but only 11,705 Feedback records (2.22\%). Ethereum contains 29,172 identities (7.54\%) and 3,209 Feedback records (0.61\%). Celo contains 9,701 identities (2.51\%) but records 27,091 Feedback entries (5.14\%), the second-largest Feedback total among the supported mainnets.

\textit{Third}, registry size does not indicate how often the EIP-8004 reputation mechanism is used. Base records 10.32 Feedback entries per registered identity, compared with 0.049 on BNB Smart Chain and 0.11 on Ethereum. Base's rate is about 210 times that of BNB Smart Chain and 94 times that of Ethereum. Moreover, four active networks contain 2,328 registered identities but no Feedback records. BNB Smart Chain leads identity deployment, whereas observed reputation activity is concentrated on Base.

\begin{observationbox}
\noindent \textbf{Observation 1.} 
EIP-8004 is widely deployed, but a large registered identity population does not necessarily correspond to active use of its reputation mechanism. BNB Smart Chain hosts 61.61\% of all identities but only 2.22\% of Feedback records, whereas Base hosts 10.50\% of the identities but 79.46\% of the Feedback records.
\end{observationbox}

\subsubsection{x402 Payment Settlement Activity}

x402 activity on Base exhibits a high-frequency, low-value payment pattern involving a relatively small buyer population and a much broader seller population. This pattern is reflected in the following two aspects.

\textit{First}, x402 processes a substantial number of settlements, but each settlement carries little monetary value. During the rolling 30-day window, the known facilitators processed 11,641,544 settlements, equivalent to an average of 388,051 settlements per day. 
These settlements transferred approximately \$639,049 in USDC in total, corresponding to about \$21,302 per day and a mean value of only \$0.055 per settlement.
Thus, x402 activity is driven primarily by payment frequency rather than by high-value transfers.

\textit{Second}, payment activity involves substantially more sellers than buyers. 
The records contain 16,892 unique buyers and 84,314 unique sellers, giving approximately five sellers for every buyer. During the 30-day window, each buyer is associated with about 689 settlements on average, while each seller receives about 138. These counts combine a high payment frequency with a broad recipient population and a mean payment of \$0.055.

\begin{observationbox}
\noindent \textbf{Observation 2.} 
x402 has established an active payment market on Base characterized by frequent, low-value settlements. The known facilitators processed 11.64 million settlements in 30 days, with a mean value of \$0.055, while the number of sellers was approximately five times the number of buyers.
\end{observationbox}

\subsubsection{MCP Service Publication and Wallet Operations}

\begin{table}[t]
\caption{MCP service publication and linked-wallet activity on Celo, Base, and BNB Smart Chain}
\label{tab:mcp_publication_wallet_activity}
\centering
\begingroup
\small
\setlength{\tabcolsep}{4.5pt}
\renewcommand{\arraystretch}{1.12}
\begin{tabular*}{\tosemtablewidth}{@{\extracolsep{\fill}}lrrrrrr@{}}
\toprule
& \multicolumn{2}{c}{\bfseries MCP publication} & \multicolumn{3}{c}{\bfseries Linked-wallet sample} & \multicolumn{1}{c}{\bfseries Transaction activity} \\
\cmidrule(lr){2-3}\cmidrule(lr){4-6}\cmidrule(l){7-7}
{\bfseries Network} & {\bfseries Publishers} & {\bfseries Share} & {\bfseries Sampled} & {\bfseries Active} & {\bfseries Active share} & {\bfseries Per active wallet} \\
\midrule
Celo & 8,578 & 88.42\% & 188 & 76 & 40.43\% & 499.3 \\
Base & 4,364 & 10.75\% & 184 & 157 & 85.33\% & 224.4 \\
BNB Smart Chain & 3,527 & 1.48\% & 233 & 212 & 90.99\% & 66.9 \\
\bottomrule
\end{tabular*}
\endgroup
\end{table}

Table~\ref{tab:mcp_publication_wallet_activity} compares MCP service publication and linked-wallet activity across the three networks with the most MCP publishers. MCP service publication is concentrated on these networks, but the on-chain activity of MCP-linked wallets varies substantially in both participation breadth and activity intensity. This pattern is reflected in three aspects: where MCP services are published, how many linked wallets are active, and what operations these wallets perform.

\textit{First}, MCP service publication is concentrated on Celo, Base, and BNB Smart Chain. Across all mainnets, 20,922 identities publish MCP services: Celo contains 8,578 (41.00\%), Base 4,364 (20.86\%), and BNB Smart Chain 3,527 (16.86\%). Together, these networks contain 78.72\% of all identities that publish MCP services. MCP publishers represent 88.42\% of registered identities on Celo, 10.75\% on Base, and 1.48\% on BNB Smart Chain. We therefore use these three networks for the detailed wallet analysis rather than selecting networks by overall registry size.

\textit{Second}, most sampled MCP service publishers are linked to wallets with observable on-chain activity, but the breadth and intensity of MCP-linked wallet activity differ across networks. Of the 605 distinct linked wallets, 445 (73.55\%) send at least one successful transaction in the observation window. The share of active wallets is 85.33\% on Base (157 of 184), 90.99\% on BNB Smart Chain (212 of 233), and 40.43\% on Celo (76 of 188).

However, a wider active-wallet population does not necessarily produce more intensive activity. The 445 active wallets send 87,359 successful transactions. Celo accounts for 43.44\% of these transactions with only 76 active wallets, or 499.3 transactions per active wallet; the corresponding values are 224.4 on Base and 66.9 on BNB Smart Chain. Base and BNB Smart Chain have a larger share of publishers connected to active wallets, whereas Celo's activity is generated by fewer but more frequently used wallets.

\textit{Third}, the identifiable activity of MCP-linked wallets is concentrated in token movement, token permissions, and exchange. DEX transactions account for 13,079 transactions (14.97\% of all transactions), ERC-20 transfers for 10,521 (12.04\%), native-token transfers for 9,780 (11.20\%), and token approvals for 8,146 (9.32\%). Lending calls account for 373 transactions (0.43\%), and transfer-from calls for 50 (0.06\%). Among the 41,949 transactions assigned a specific type, token transfers and approvals account for 67.81\%, while DEX transactions account for 31.18\%. The remaining 45,410 transactions (51.98\%) are unclassified contract interactions because many custom or proxy calls cannot be mapped to one operation with confidence.

\begin{observationbox}
\noindent \textbf{Observation 3.} 
MCP publication and wallet activity follow different patterns across networks. Celo contains 41.00\% of all identities that publish MCP services, while 445 of the 605 linked wallets in the three-network sample send 87,359 successful transactions. Celo has the lowest share of active wallets (40.43\%) but the highest number of transactions per active wallet (499.3). Publishing an MCP service is therefore associated with observable on-chain use, but the breadth and intensity of participation differ across networks.
\end{observationbox}

\begin{answerbox}
\noindent \textbf{Answer to RQ1:}
The Web4 agent economy is fragmented across layers and networks: identity registration, reputation use, and MCP publication are not synchronized; x402 settlements reveal a high-throughput, low-value service market; and MCP publication does not align with the breadth or intensity of linked-wallet activity.
\end{answerbox}

\section{RQ2: Challenges in Web4 Agent Development}
\label{sec:04_RQ2_Imbalances&Risks}
RQ1 shows that Web4 already supports registered identities, machine payments, and wallet activity. However, scale does not show whether developers can build and operate these systems reliably. RQ2 therefore examines concrete engineering problems reported in Web4-related GitHub repositories.

\subsection{Methodology}

\begin{figure}[t]
\centering
\includegraphics[page=2,width=\textwidth]{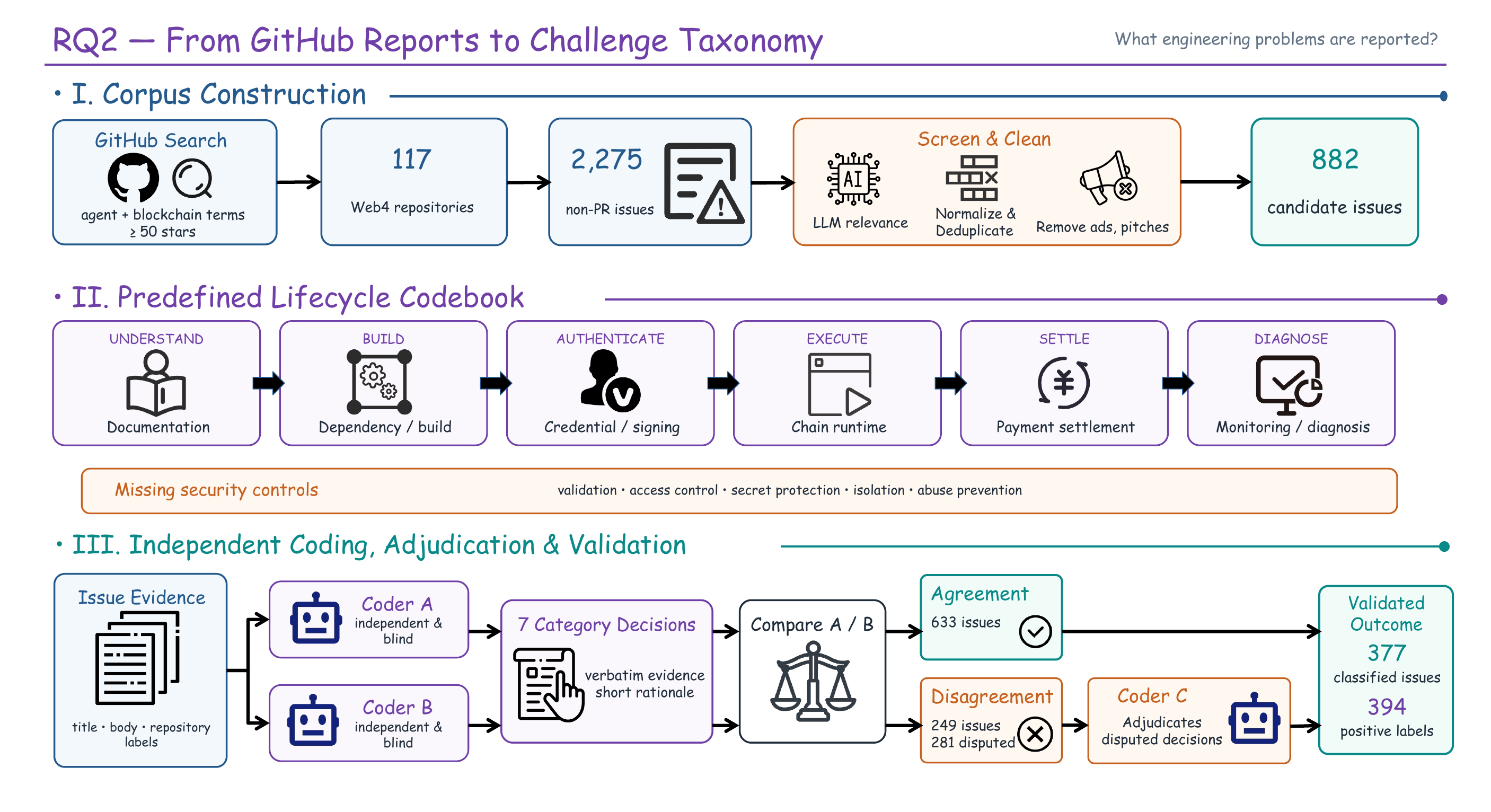}
\caption{Overview of corpus construction, lifecycle-based coding, independent classification, and adjudication for RQ2.}
\Description{The RQ2 pipeline searches GitHub to identify 117 Web4 repositories and 2,275 non-pull-request issues, screens them to 882 candidate issues, applies a seven-category lifecycle codebook with missing security controls as a cross-cutting category, and uses two independent coders plus adjudication to produce 377 classified issues with 394 labels.}
\label{fig:rq2-methodology-overview}
\end{figure}

\noindent \textbf{Overview.} We construct a corpus of Web4-related issues, classify them into seven predefined categories, and resolve disagreements between two independent LLMs through separate adjudication. Figure~\ref{fig:rq2-methodology-overview} summarizes the corpus construction, independent coding, and adjudication process. Corpus filtering identifies relevant problem reports, while evidence-based coding makes each assigned category traceable to the issue text. We analyze the category results to identify engineering problems in Web4 development.

\subsubsection{Repository and Issue Corpus}

We build a corpus of public GitHub issues that report concrete problems in Web4 development.

\noindent \textbf{Project inventory.} We first built the Web4 project inventory through GitHub repository search. We paired agent terms (\textit{Web4}, \textit{AI agent}, \textit{autonomous agent}, or \textit{MCP}) with blockchain terms such as \textit{Web3}, \textit{smart contract}, \textit{wallet}, and \textit{payment}. Requiring at least 50 stars and deduplicating results across queries produced 117 repositories.

\noindent \textbf{Issue collection.} We queried the GitHub REST API on 16 June 2026 for these repositories and collected non-pull-request issues updated since 1 June 2025~\cite{github2026restapi}. GitHub's issue endpoint also returns pull requests, so we removed those records to keep the unit of analysis consistent. In total, the 117 repositories returned 2,275 issues. We retained each issue's title, body, and repository labels because these fields describe the problem as submitted, before later comments or fixes can change its interpretation.

\noindent \textbf{Relevance filtering and deduplication.} We first retained issues that directly mentioned a Web4 protocol and described an engineering problem. We then added issues that described related credential, chain, security, payment, or diagnostic problems, even without terms such as \textit{MCP} or \textit{x402}. This second step prevents relevant reports from being missed. To ensure that the corpus contains concrete Web4 engineering problems, we also use an LLM to remove advertisements, catalog requests, integration pitches, feature requests without a current obstacle, unrelated application issues, and records with insufficient information. These steps produced 882 candidate issues.

\subsubsection{Seven Categories of practical challenges.}

We define seven categories to organize the practical problems reported across the Web4 development process.

\noindent \textbf{Category design.} We define the seven categories listed in Table~\ref{tab:rq2_categories}. The categories follow the Web4 development lifecycle: understand and configure, build and deploy, authenticate and sign, execute on-chain, settle payment, and diagnose operation. Security is cross-cutting because a missing safeguard can occur at any stage. This design follows lifecycle-oriented defect classification~\cite{chillarege1992odc}; prior work also treats documentation, build, and logging problems as distinct engineering concerns~\cite{aghajani2019documentation,seo2014build,yuan2012logging}, while the NIST SSDF treats security as a lifecycle-wide activity~\cite{nist2022ssdf}. Each category has an operational boundary that specifies the problems it includes~\ref{tab:rq2_categories}.

\begin{table}[t]
\caption{Seven categories of practical challenges}
\label{tab:rq2_categories}
\centering
\begingroup
\small
\setlength{\tabcolsep}{5.5pt}
\renewcommand{\arraystretch}{1.10}
\begin{tabular*}{\tosemtablewidth}{@{\extracolsep{\fill}}p{0.30\textwidth}p{0.615\textwidth}@{}}
\toprule
{\bfseries Category} & {\bfseries Operational boundary} \\
\midrule
Credential and signing failures & Failed credential setup, authentication, authorization, wallet sessions, or signature verification. \\
Chain runtime failures & RPC, gas, nonce, chain-state, transaction, contract-call, or network-specific execution failures. \\
Documentation gaps & Missing, incorrect, outdated, or non-actionable setup and usage information. \\
Dependency and build failures & Package resolution, installation, compilation, container, module, or CI failures. \\
Missing security controls & Missing validation, access control, secret protection, isolation, rate limiting, or abuse prevention. \\
Payment settlement failures & HTTP 402/x402 authorization, verification, settlement, token, amount, or ordering failures. \\
Monitoring and diagnosis gaps & Missing logs, metrics, traces, health checks, correlation identifiers, or actionable error context. \\
\bottomrule
\end{tabular*}
\endgroup
\end{table}

\noindent \textbf{Multi-label rule.} The scheme was multi-label because one report can contain two separate problems, such as a runtime failure and missing diagnostic context~\cite{kim2021multilabel}. A second label was allowed only when it had independent textual evidence; this rule preserves genuine overlap without adding categories from expected co-occurrence. An issue with no supported category was marked unclassified rather than forced into the closest category.

\subsubsection{Independent Coding and Adjudication}

We use independent classification and a separate review step to make the issue labels more reliable.

\noindent \textbf{Independent evidence-based coding.} Two LLMs independently evaluated all seven categories for every candidate issue. Retrieval keywords, previous labels, the other LLM's output, and final category counts were hidden from both LLMs. Hiding these signals reduces bias from the retrieval process or an expected distribution. Each positive decision required a supporting excerpt from the title, body, or repository labels and a short explanation linking that excerpt to the corresponding category definition. This evidence requirement makes every positive label traceable.

\noindent \textbf{Agreement measurement.} Table~\ref{tab:rq2_agreement} reports the positive labels assigned by each LLM. A+ and B+ are the numbers of LLM A and LLM B before adjudication, Both+ is their positive overlap, and Final is the adjudicated count. In total, 281 decisions from 249 issues required adjudication. A third LLM received the issue text, category definition, and competing decisions. Separate adjudication was used because accepting one LLM by default would preserve that LLM's systematic boundary errors. The third LLM retained a label only when the issue text satisfied its operational boundary.

\begin{table}[t]
\caption{Positive labels before and after adjudication}
\label{tab:rq2_agreement}
\centering
\begingroup
\small
\setlength{\tabcolsep}{5.5pt}
\renewcommand{\arraystretch}{1.10}
\begin{tabular*}{\tosemtablewidth}{@{\extracolsep{\fill}}lrrrrr@{}}
\toprule
& \multicolumn{3}{c}{\bfseries Before adjudication} & \multicolumn{1}{c}{\bfseries After} & \multicolumn{1}{c}{\bfseries Agreement} \\
\cmidrule(lr){2-4}\cmidrule(lr){5-5}\cmidrule(l){6-6}
{\bfseries Category} & {\bfseries A+} & {\bfseries B+} & {\bfseries Both+} & {\bfseries Final} & {\bfseries $\kappa$} \\
\midrule
Credential/signing & 32 & 32 & 25 & 33 & 0.773 \\
Chain runtime & 19 & 113 & 17 & 22 & 0.229 \\
Documentation & 54 & 71 & 39 & 72 & 0.596 \\
Dependency/build & 75 & 65 & 58 & 72 & 0.814 \\
Security controls & 107 & 103 & 80 & 118 & 0.730 \\
Payment settlement & 9 & 9 & 7 & 9 & 0.775 \\
Monitoring/diagnosis & 60 & 42 & 29 & 68 & 0.543 \\
\bottomrule
\end{tabular*}
\endgroup
\end{table}

\noindent \textbf{Disagreement resolution.} The largest disagreement concerned chain runtime failures. LLM A assigned 19 positives and LLM B assigned 113; they shared 17 positives, leaving 98 disputed decisions. LLM B included ordinary API errors, model-call failures, and general process crashes, whereas the category definition requires a direct link to RPC behavior, gas, nonce, chain state, transaction submission, contract invocation, or network-specific execution. The third LLM retained five disputed positives, producing 22 final chain-runtime issues. Documentation disagreements mainly concerned whether a question demonstrated missing guidance, while monitoring disagreements concerned whether logs were absent rather than merely quoted in the report. Other categories have maintained a relatively high level of consistency between the two LLMs.

\subsection{Distribution of Reported Challenges}

The coding process identifies 377 issues that contain evidence for at least one predefined challenge category. These issues cover 75 repositories and contain 394 positive labels: 360 issues have one label and 17 have two. The other 505 of the 882 candidates may discuss Web4 projects, but their text does not support any of the seven engineering categories, so they are excluded from category-frequency analysis.

\begin{table}[t]
\caption{Reported Web4 development challenges (multi-label)}
\label{tab:rq2_concrete_themes}
\centering
\begingroup
\small
\setlength{\tabcolsep}{5.5pt}
\renewcommand{\arraystretch}{1.10}
\begin{tabular*}{\tosemtablewidth}{@{\extracolsep{\fill}}lrr@{}}
\toprule
{\bfseries Category} & {\bfseries Issues} & {\bfseries Share of 377} \\
\midrule
Missing security controls & 118 & 31.3\% \\
Documentation gaps & 72 & 19.1\% \\
Dependency and build failures & 72 & 19.1\% \\
Monitoring and diagnosis gaps & 68 & 18.0\% \\
Credential and signing failures & 33 & 8.8\% \\
Chain runtime failures & 22 & 5.8\% \\
Payment settlement failures & 9 & 2.4\% \\
\bottomrule
\end{tabular*}
\endgroup
\end{table}

\noindent \textbf{Missing security controls.} Missing safeguards allow agent-selected inputs to reach the operating system, network, or secrets without an enforced restriction. This is the largest category, with 118 issues (31.3\%); the reports identify absent safeguards rather than requiring evidence of a completed attack. For example, an MCP execution tool can run arbitrary shell commands without a sandbox, allowlist, or input validation~\cite{teleton2026permissions}; an unvalidated network parameter can enable server-side request forgery and API-key exfiltration~\cite{alchemy2026ssrf}; and another server writes API credentials to INFO logs~\cite{polymarket2026credlog}. In Web4 systems, these controls must be applied before tool execution because the agent may invoke tools without a human checking each argument.

\noindent \textbf{Documentation gaps.} Incorrect instructions prevent users from correcting failures, while missing guidance prevents them from reaching a runnable configuration. The 72 documentation issues (19.1\%) cover both forms. For example, a typo in the Foundry MCP setup command returns only \texttt{Invalid input}, without identifying the incorrect field~\cite{foundry2026unclearerror}. Other reports request onboarding instructions for SDK, MCP, and npm components~\cite{omnuron2026onboarding}.

\noindent \textbf{Dependency and build failures.} These failures block installation or startup before a Web4 function can run. The 72 issues (19.1\%) include incompatible package constraints, missing runtime modules, and unavailable package versions. For example, one MCP server cannot resolve conflicting FastAPI, MCP, and \texttt{anyio} requirements~\cite{polymarket2026depconflict}; another fails because \texttt{pytz} is absent~\cite{alpaca2026nomodule}, and its Docker build requests an \texttt{httpcore} version that cannot be downloaded~\cite{alpaca2026dockerbuild}.

\noindent \textbf{Monitoring and diagnosis gaps.} Missing identifiers and current metrics prevent operators from locating whether a failure began in the agent request, signer, RPC provider, or transaction path. The 68 issues in this category (18.0\%) concern missing or misleading evidence about system state. One report requests correlation identifiers that connect an MCP request to signer activity and transaction execution~\cite{starknet2026auditlogs}; another requests per-tool usage, latency, error, and token metrics~\cite{rimuru2026observability}. The stale-data report in \texttt{polymarket-mcp-server}\#2 was assigned both chain-runtime and monitoring labels because it returns outdated market state and provides no reliable view of current state~\cite{polymarket2026staledata}.

\begin{figure}[t]
\centering
\begin{issuebox}
\textbf{Title:} \texttt{gmgn-cli order strategy list} returns \texttt{401 AUTH\_INVALID: missing signature}\\[0.4em]
\textbf{Command:} \texttt{gmgn-cli order strategy list --chain sol}\\[0.3em]
\textbf{Observed:} The request returns HTTP 401 with \texttt{AUTH\_INVALID: missing signature}.\\[0.3em]
\textbf{Expected:} Return the open strategy orders for the wallet bound to the API key.\\[0.3em]
\textbf{Reported cause:} The help text describes normal authentication, but the endpoint requires a signed request.\\[0.3em]
\textbf{RQ2 category:} Credential and signing failures.
\end{issuebox}
\caption{A representative issue showing a missing request signature (\texttt{GMGNAI/gmgn-skills}\#72~\cite{gmgn2026signature}).}
\Description{A compact issue summary showing the affected command, an HTTP 401 missing-signature error, the expected order-list response, the reported authentication mismatch, and its assignment to the credential and signing failures category.}
\label{fig:rq2_issue_example}
\end{figure}

\noindent \textbf{Credential and signing failures.} Delegated actions fail before execution when credentials are unavailable to the client or a protected request is not signed correctly. The 33 issues (8.8\%) cover these two failure points. The Alpaca MCP Docker image returns \texttt{request is not authorized} despite configured API keys~\cite{alpaca2026unauthorized}. Figure~\ref{fig:rq2_issue_example} shows the second failure point: a CLI calls a protected endpoint without the required signature. The first case concerns credential availability in the runtime environment, whereas the second concerns protocol-specific request signing.

The example provides the information needed to verify its category assignment. Figure~\ref{fig:rq2_issue_example} states the command, observed error, expected behavior, and reported cause in a compact form.

\noindent \textbf{Chain runtime failures.} These failures either expose an agent to stale state or prevent its transaction from being submitted. The 22 chain-runtime issues (5.8\%) affect state retrieval or transaction submission. \texttt{polymarket-mcp-server}\#2 returns old market data rather than the current market set~\cite{polymarket2026staledata}. In \texttt{goat-sdk/goat}\#515, an approval transaction passes a hexadecimal string where \texttt{walletClient.sendTransaction} requires a \texttt{bigint}, so submission fails~\cite{goat2026transaction}.

\noindent \textbf{Payment settlement failures.} Incorrect ordering or enforcement of x402 operations can deliver work without payment or reject a valid payment before delivery. The nine payment issues (2.4\%) include three failure modes. \nolinkurl{GetBindu/Bindu}\#562 persists completed artifacts after settlement fails~\cite{bindu2026settlementgate}; \nolinkurl{nirholas/agenti}\#111 lets a paid handler run after verification without completing settlement~\cite{agenti2026settlementgate}; and another report records an authorized x402 payment rejected by the server during verification~\cite{agentkit2026x402paypath}. These cases represent work delivered without settlement, verification treated as settlement, and valid payment rejected before delivery.

Most reported challenges concern the controls and engineering support needed to operate Web4 software. Security controls, documentation, dependency/build, and monitoring account for 330 of the 394 positive labels (83.8\%). The remaining 64 labels (16.2\%) concern credential/signing, chain runtime, or payment settlement and occur at Web4-specific authorization, transaction, and settlement boundaries.

\begin{observationbox}
\noindent \textbf{Observation 4.} Of 394 positive labels, 330 (83.8\%) concern security controls, documentation, dependency/build, or monitoring. These issues prevent safe tool use, reproducible setup, successful startup, or fault localization. The other 64 labels (16.2\%) occur at Web4-specific authorization, chain-execution, and payment-settlement boundaries.
\end{observationbox}

\begin{answerbox}
\noindent \textbf{Answer to RQ2:}
Web4 development is mainly hindered by missing security controls and insufficient documentation, build support, and diagnostic information. Additional failures occur at the Web4-specific boundaries of delegated authorization, on-chain execution, and payment settlement.
\end{answerbox}

\section{RQ3: Response Practices in Web4 Development}
\label{sec:05_RQ3_FutureChallenge}
RQ2 identifies the engineering problems reported in Web4 repositories. RQ3 asks a different question: what observable action did the project take after each report? We separate problem reports from project responses because an issue body describes the reporter's claim, whereas comments, pull requests, commits, and file changes show how the project reacted.

\subsection{Methodology}

\begin{figure}[t]
\centering
\includegraphics[page=3,width=\textwidth]{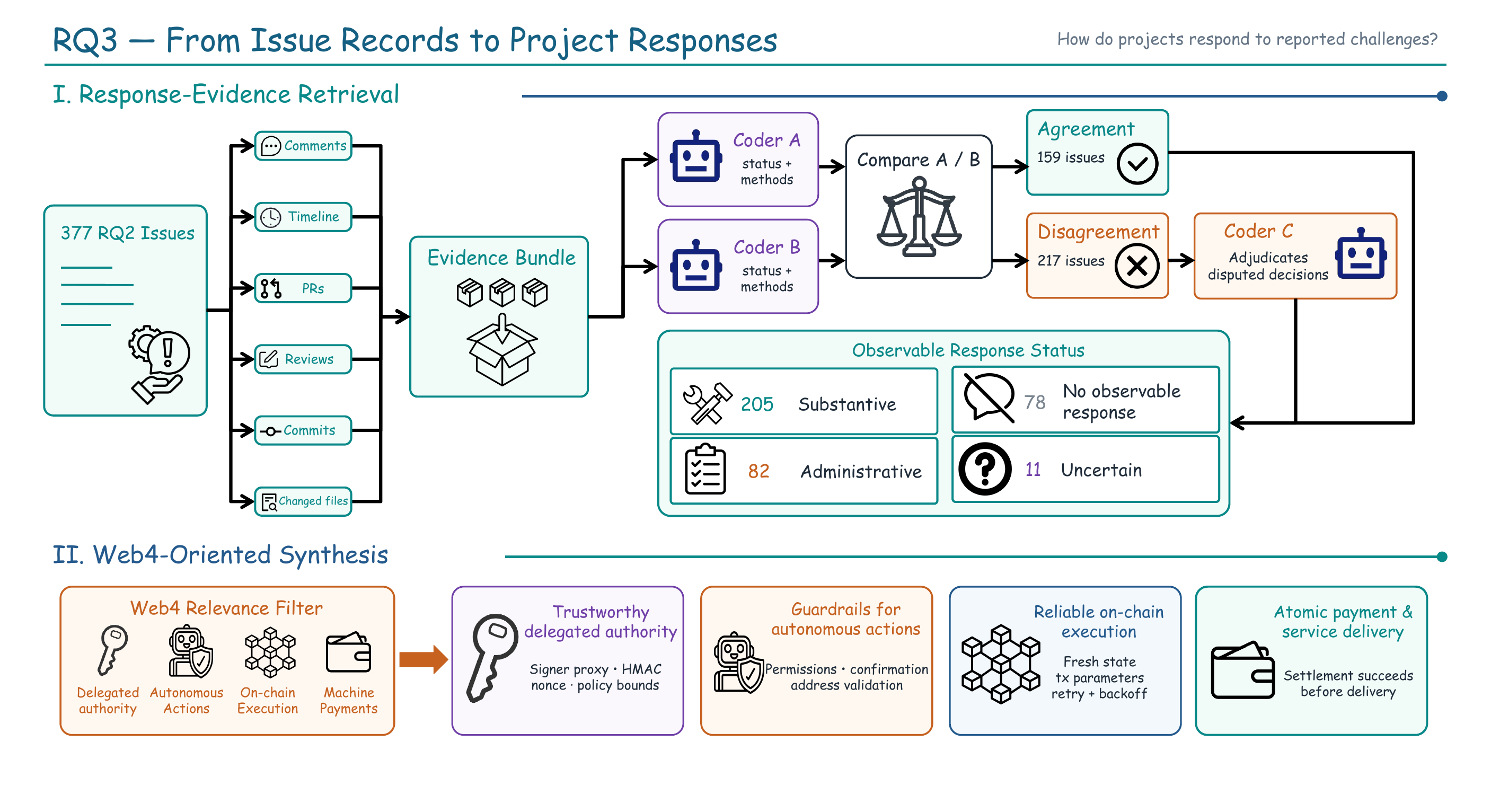}
\caption{Overview of response-evidence retrieval, independent coding, adjudication, validation, and Web4-oriented synthesis for RQ3.}
\Description{The RQ3 pipeline retrieves comments, timelines, pull requests, reviews, commits, and changed files for 377 issues; packages the evidence for two independent coders and adjudication; reports 205 substantive, 82 administrative, 78 no-observable-response, and 11 uncertain outcomes; and synthesizes four Web4 remediation patterns covering delegated authority, autonomous-action guardrails, on-chain execution, and payment-before-delivery.}
\label{fig:rq3-methodology-overview}
\end{figure}

\noindent \textbf{Overview.} RQ3 starts from the issues classified in RQ2, collects their project-response evidence, codes response status and methods, and synthesizes actions that depend on Web4-specific properties. Figure~\ref{fig:rq3-methodology-overview} summarizes how this evidence is collected, coded, validated, and synthesized. Separating evidence collection, coding, and synthesis prevents an issue's reported problem from being treated as evidence of a project response.

\subsubsection{Response-Evidence Corpus}

We collect project actions linked to each RQ2 issue to determine how maintainers responded.

\noindent \textbf{Evidence retrieval.} RQ3 starts from the 377 issues classified in RQ2 and keeps their RQ2 labels unchanged. On 7 August 2026, we retrieved issue comments, timeline events, linked pull requests, pull-request discussions and reviews, commits, and changed-file lists through the GitHub API. Comments capture maintainer explanations and usage guidance; timeline events capture assignment, transfer, duplicate, and closure actions; pull requests, commits, and file lists provide evidence of implemented code, documentation, configuration, and test changes. One issue returned HTTP 404 and was excluded rather than counted as no response, leaving 376 analyzable issues.

\noindent \textbf{Evidence-bundle construction.} We constructed one normalized evidence bundle for each issue. A bundle contained the issue title, body, state, labels, and assignees; all retrieved issue comments with author roles, timestamps, and URLs; relevant timeline events; and each linked pull request's merge status, changed-file names, commit messages, discussions, and reviews. The issue title and body provided problem context, while the remaining fields showed how the project responded.

\subsubsection{Response Coding Scheme}

We classify both the response outcome and the methods used to address each issue.

\noindent \textbf{Response status.} We first assigned one mutually exclusive response status to each issue:
\begin{itemize}
    \item \textit{substantive response}: an accepted technical change, documentation change, test, configuration change, diagnosis, or actionable maintainer guidance;
    \item \textit{administrative only}: assignment, labeling, transfer, duplicate marking, closure, or scope management without a technical action;
    \item \textit{no observable response}: no project-side technical or administrative action in the retrieved record; and
    \item \textit{uncertain}: available evidence could not support a stable decision.
\end{itemize}
We keep administrative responses separate because closing or labeling an issue does not show that the reported problem was repaired~\cite{kalliamvakou2014promises}. We also keep uncertainty as an explicit outcome so that incomplete evidence is not forced into a positive or negative status.

\noindent \textbf{Response methods.} Before coding, we defined a study-specific response-method codebook. Drawing on software-maintenance research, we classified observable actions and changed artifacts into implementation fixes, configuration or dependency changes, documentation or example updates, usage guidance, and testing or validation~\cite{iso14764maintenance,chapin2001maintenance}. We added security hardening, diagnostic improvement, and protocol or infrastructure adjustment to cover the security, observability, signer, RPC, chain, and payment boundaries examined in RQ2~\cite{nist2022ssdf,yuan2012logging}. Triage and no substantive response were kept separate so that labeling, assignment, or closure was not counted as a repair~\cite{kalliamvakou2014promises}. Methods were multi-label because one response could change code, tests, and documentation together.

\noindent \textbf{Evidence requirements.} The coders were instructed not to treat an RQ2 category, the issue body alone, a bot message, an unaccepted external pull request, or the issue's closed state as evidence of a project response. Every positive response method required an action summary, actor type, source URL, and excerpt from the bundle. These rules require each positive decision to point to an observable project action and prevent a proposal or closure event from being counted as an implemented response.

\subsubsection{Independent Coding and Validation}

We independently code each response and check that every decision is supported by the collected evidence.

\noindent \textbf{Independent coding and adjudication.} Two LLM coders independently applied the same status and method codebook to the normalized evidence bundles. The coders agreed on response status for 286 of 376 issues (76.1\%, Cohen's $\kappa=0.580$). They assigned the same complete method set to 175 issues (46.5\%), with mean Jaccard similarity 0.603. Exact method agreement is lower because one coder may record the code change while the other also records its tests or documentation. The coders agreed on both status and complete method set for 159 issues; a third LLM adjudicator reviewed the evidence and competing decisions for the remaining 217.

\noindent \textbf{Agreement and traceability checks.} Mechanical validation confirmed 376 unique issue identifiers, 503 method assignments with non-empty evidence, and 802 evidence URLs present in the source corpus. These checks ensure that each coded action can be traced to the retrieved evidence.

\subsubsection{Web4-Oriented Analysis}

We identify the responses that address problems specific to Web4 systems.

\noindent \textbf{Web4 relevance criteria.} The response-method codebook also includes ordinary software maintenance, such as package upgrades and README corrections. To keep the final synthesis aligned with Web4, we retained adjudicated action summaries only when the action depended on at least one of four properties: delegated agent authority, autonomous tool execution, on-chain state or transactions, or machine-payment settlement. This criterion prevents a generic build fix from being presented as a Web4-specific contribution.

\subsection{Observed Project Responses}

Table~\ref{tab:rq3_status} reports the response statuses. A substantive response was visible for 205 issues (54.5\%). The other records contained only administrative action (82, 21.8\%), no observable response (78, 20.7\%), or evidence that remained uncertain after adjudication (11, 2.9\%). 

\begin{table}[t]
\caption{Observable response status for 376 RQ2 issues}
\label{tab:rq3_status}
\centering
\begingroup
\small
\setlength{\tabcolsep}{5.5pt}
\renewcommand{\arraystretch}{1.10}
\begin{tabular*}{\tosemtablewidth}{@{\extracolsep{\fill}}lrr@{}}
\toprule
{\bfseries Response status} & {\bfseries Issues} & {\bfseries Share} \\
\midrule
Substantive response & 205 & 54.5\% \\
Administrative only & 82 & 21.8\% \\
No observable response & 78 & 20.7\% \\
Uncertain & 11 & 2.9\% \\
\bottomrule
\end{tabular*}
\endgroup
\end{table}

Substantive responses appear more often in dependency/build and monitoring issues than in chain-runtime and payment-settlement issues. Table~\ref{tab:rq3_status_by_category} records substantive responses for 52 of 72 dependency/build issues and 42 of 68 monitoring issues, compared with 7 of 22 chain-runtime issues and 1 of 9 payment-settlement issues. The security category contains 55 substantive and 36 administrative responses.

\begin{table}[t]
\caption{Response status by RQ2 category (RQ2 labels may overlap)}
\label{tab:rq3_status_by_category}
\centering
\begingroup
\small
\setlength{\tabcolsep}{5.5pt}
\renewcommand{\arraystretch}{1.10}
\begin{tabular*}{\tosemtablewidth}{@{\extracolsep{\fill}}lrrrrr@{}}
\toprule
& & \multicolumn{4}{c}{\bfseries Response status} \\
\cmidrule(l){3-6}
{\bfseries RQ2 category} & {\bfseries $N$} & {\bfseries Subst.} & {\bfseries Admin.} & {\bfseries None} & {\bfseries Uncert.} \\
\midrule
Credential/signing & 33 & 19 & 6 & 6 & 2 \\
Chain runtime & 22 & 7 & 6 & 9 & 0 \\
Documentation & 72 & 39 & 12 & 17 & 4 \\
Dependency/build & 72 & 52 & 5 & 14 & 1 \\
Security controls & 117 & 55 & 36 & 24 & 2 \\
Payment settlement & 9 & 1 & 1 & 6 & 1 \\
Monitoring/diagnosis & 68 & 42 & 19 & 6 & 1 \\
\bottomrule
\end{tabular*}
\endgroup
\end{table}

Projects respond most through both issue management and technical changes. Among the 503 response-method assignments in Table~\ref{tab:rq3_methods}, triage or scope management appears 92 times, followed by implementation fixes (89), testing or validation (69), documentation updates (49), and security hardening (44). Method totals exceed 376 because one issue can contain several distinct actions.

\begin{table}[t]
\caption{Observable response methods (multi-label)}
\label{tab:rq3_methods}
\centering
\begingroup
\small
\setlength{\tabcolsep}{5.5pt}
\renewcommand{\arraystretch}{1.10}
\begin{tabular*}{\tosemtablewidth}{@{\extracolsep{\fill}}lrr@{}}
\toprule
{\bfseries Response method} & {\bfseries Assignments} & {\bfseries Share of 376} \\
\midrule
Triage or scope management & 92 & 24.5\% \\
Implementation fix & 89 & 23.7\% \\
Testing or validation & 69 & 18.4\% \\
Documentation or example update & 49 & 13.0\% \\
Security or policy hardening & 44 & 11.7\% \\
Workaround or usage guidance & 35 & 9.3\% \\
Diagnostic or monitoring improvement & 24 & 6.4\% \\
Configuration or dependency change & 17 & 4.5\% \\
Protocol or infrastructure adjustment & 6 & 1.6\% \\
No substantive response & 78 & 20.7\% \\
\bottomrule
\end{tabular*}
\endgroup
\end{table}

\begin{observationbox}
\noindent \textbf{Observation 5.} Of 376 accessible issues, 205 received a substantive response, while 82 received only administrative action and 78 had no observable project response. Repository activity therefore does not by itself demonstrate a technical remedy.
\end{observationbox}

\subsection{Web4-Oriented Remediation Patterns}

We next focus on how projects respond to Web4-specific issues. These responses form four patterns: securing delegated authority, constraining autonomous actions, improving on-chain execution, and enforcing payment before service delivery.

\textbf{Trustworthy delegated authority.} Web4 projects protect delegated authority by separating signing from application logic and checking which requests may be signed. In \nolinkurl{GMGNAI/gmgn-skills}\#72, the order-list command used normal authentication and sent no signature, so the protected endpoint returned a 401 error. The merged fix switched the command to signature-based authentication~\cite{gmgn2026signature}. In \texttt{starknet-agentic}\#214, signing and application logic were not separated, so an attack on the application could also expose signing authority. The project moved signing to a dedicated proxy with request authentication, replay protection, and policy and chain checks~\cite{starknet2026trustboundary}.

\textbf{Safety controls for autonomous actions.} Projects add permission, confirmation, and destination checks before sensitive actions. Teleton Agent issue \#17 allowed its ``yolo'' mode to run any shell command without a sandbox, allowlist, or input checks. The project added access controls for individual tools~\cite{teleton2026permissions}. SeekerClaw issue \#207 allowed the agent to use the camera and location without user confirmation and accepted some invalid Solana addresses. The fix required confirmation for these tools and strengthened address validation~\cite{seekerclaw2026confirmation}.

\textbf{Reliable on-chain execution.} Projects repair incorrect chain data, invalid transaction inputs, and long waits after network failures. Polymarket MCP issue \#2 returned closed markets from 2020--2021 instead of current data. The fix excluded closed and expired markets and added tests~\cite{polymarket2026staledata}. In GOAT issue \#515, an approval transaction passed a hexadecimal string where the wallet client required an integer, so submission failed. The fix corrected the value type~\cite{goat2026transaction}. In Valory Trader issue \#938, failed API calls could block the only worker for long periods, and stale metrics lacked an update time. The fix added shorter retry delays, exposed the last update time, and added tests~\cite{valory2026resilience}.

\textbf{Payment before service delivery.} Payment must succeed before a paid task runs or delivers its output. In Bindu issue \#562, the system could mark a task as complete and store its output even when x402 settlement failed. The merged changes moved settlement before task execution, blocked output delivery after a failed settlement, and added unit and end-to-end tests~\cite{bindu2026settlementgate}.

\begin{observationbox}
\noindent \textbf{Observation 6.} Evidence-backed Web4 responses establish signing boundaries, add confirmation and permission controls, repair chain-facing state and transaction handling, and make settlement a prerequisite for service delivery.
\end{observationbox}

\begin{answerbox}
\noindent \textbf{Answer to RQ3:}
Projects address general engineering issues through enhancing code, configuration, tests, documentation, and diagnostics. For Web4-specific issues, they isolate signing and enforce permissions, filter stale chain state, fix transaction inputs and retry temporary failures, and deliver services only after payment settles
\end{answerbox}

\section{Discussion}
\label{sec:06_Discussion}
The three RQs show that Web4 has moved from protocol design to visible use. Identity registration, reputation, machine payments, and MCP-linked wallet activity are all present at scale, while repository evidence shows how projects are improving the software behind them. These findings lead to four practical implications.

\subsection{Implications}

\subsubsection{Build a Checked Web4 Workflow}

For requests that can move assets or change on-chain state, developers should use a fixed sequence: authenticate the request, confirm that the tool is allowed, validate the network, address, and amount, sign through a separate service, confirm any required transaction or payment, and only then return the result. Each step should record the same request identifier, the permission decision, transaction hash, and payment status so that operators can trace and audit the action.

If any check, RPC call, signature, transaction, or settlement fails or times out, the system should stop instead of using stale data or delivering an unpaid result. Before release, teams should test invalid addresses, repeated authorizations, stale chain data, RPC timeouts, reverted transactions, and failed settlements. These cases directly cover the failures observed in RQ2 and the repair patterns identified in RQ3.

\subsubsection{Make Agent Actions Safe by Default}

RQ2 shows that missing security controls are the most common reported challenge. RQ3 also shows that projects already use several practical fixes. They place signing behind an authenticated service, limit requests with policies and nonces, and add permission checks, confirmation gates, and address validation.

These controls should become standard parts of agent systems. Agent reasoning should be kept separate from signing keys, and each tool should receive only the permissions it needs. Before a sensitive action, the system should check the destination address, network, amount, and user intent. Agent frameworks can provide these checks by default and apply common secure-development guidance~\cite{nist2022ssdf,owasp2023top10llm} directly to tool calls and transactions.

\subsubsection{Test the Complete Service Flow}

Web4 services depend on several connected parts: agent software, wallets, blockchain nodes, smart contracts, and payment services. RQ2 shows that documentation, build, and monitoring problems can stop a service before an on-chain action begins. Chain-state, transaction, and payment errors can then interrupt later steps.

RQ3 identifies useful ways to make this flow more reliable. Projects filter stale chain state, correct transaction parameters, retry temporary RPC failures, and add tests for failure paths. For x402 services, the required order is clear: verify the payment, complete settlement, and then deliver the service. Projects should test both successful and failed cases at every step and record which step caused an error. Shared test suites and clear error messages would make these checks easier to reuse across projects.

\subsubsection{Turn Project Fixes into Shared Practice}

Among the 376 analyzed issues, 205 received a substantive response. These responses often combine code changes with tests, documentation, usage guidance, security controls, or better diagnostics. Together, they provide practical examples of how Web4 projects solve recurring development problems.

Maintainers can make these solutions easier to reuse by linking issues to merged changes, adding regression tests, and recording the final fix when an issue is closed. This also gives clearer evidence of maintenance work than issue status alone~\cite{kalliamvakou2014promises}. Tool builders can turn common fixes into templates, checklists, default security settings, and shared test cases. In this way, solutions developed in one repository can support the wider Web4 ecosystem.

\subsection{Threats to Validity}

\subsubsection{External Threats}

Our study covers EIP-8004 records on supported mainnets, Base USDC settlements involving known x402 facilitators, sampled MCP-linked wallets on three networks, and selected public GitHub repositories; activity outside these sources may follow different patterns. We reduce this risk by using multi-chain registry data, explicit selection and measurement rules, and a broad set of public repositories. Our conclusions therefore describe the measured ecosystem and time period rather than every Web4 system, and future measurements may produce different results.

\subsubsection{Internal Threats}

Measurement and coding errors may affect our results: facilitator-based rules may miss some x402 settlements, a transaction from an MCP-linked wallet does not prove that an agent initiated it, function selectors may not reveal custom or proxy calls, and LLM-based issue coding may be incorrect. We reduce this risk by matching facilitator calls with USDC transfer events, analyzing only successful transactions sent by linked wallets, leaving uncertain operations unclassified, using two independent LLMs and a third LLM to resolve disagreements, and requiring supporting evidence for each issue label. These checks improve traceability but cannot remove all errors; our results represent observable associated activity and reported development issues rather than complete causal attribution or prevalence.

\section{Related Work}
\label{sec:07_RelatedWork}

The evolution toward Web4 lies at the intersection of Web3 empirical research and the development of Large Language Model (LLM) based autonomous agents. Our work builds upon and extends the existing literature in both domains.

\subsection{Empirical Studies on Web3 and Smart Contracts} 

Historically, empirical Web3 research has predominantly focused on the security and ecosystem dynamics of human-driven smart contracts \cite{luu2016making,victor2019measuring}. Chen et al. constructed money-transfer, contract-creation, and contract-invocation graphs to characterize Ethereum activities and relationships among accounts and contracts~\cite{chen2018ethereumgraph}. These studies laid crucial groundwork for on-chain analytics but remain confined to a ``human-execution'' paradigm. Our research shifts this focus by providing the large-scale empirical mapping of \textit{agent-driven} on-chain behaviors within the emerging Machine-to-Machine (M2M) economy.

\subsection{LLM-based Autonomous Agents and Tool Invocation} 

The rapid advancement of LLMs has catalyzed extensive research into autonomous agents \cite{yao2022react,wang2024survey}. However, existing literature primarily evaluates these agents in isolated, off-chain Web2 environments (e.g., standard APIs or closed sandboxes). Our study bridges this gap by empirically diagnosing agent deployment in permissionless Web3 environments, revealing unique engineering challenges—such as the lack of secure on-chain sandboxes—when agents utilize protocols like MCP to interact with high-risk assets.

\subsection{Decentralized Identity and M2M Economies} 

While M2M economies and decentralized identifiers (DIDs) are widely discussed in theoretical frameworks~\cite{christidis2016blockchains,sporny2022did,kobayashi2026web4}, most works lack empirical validation in large-scale production environments~\cite{nath2015comes,zhou2023review,gurpinar2025web4,palatov2025web4}. By triangulating multi-chain EIP-8004 registrations~\cite{eip8004trustless} with x402 micro-payment logs~\cite{coinbase2025x402,x402scan2026index}, our work transitions the discourse from theoretical design to quantitative engineering reality, highlighting the structural imbalances and the critical ``Validation Chasm'' that impede true trustlessness.

\section{Conclusion}
\label{sec:08_Conclusion}
This paper presents a large-scale empirical study of the Web4 agent ecosystem by examining EIP-8004 registry and reputation data, x402 settlements, the on-chain activity of MCP-linked wallets, and developer-reported issues and project responses. Our findings show that Web4 has moved beyond a conceptual vision: agents register identities, publish services, make frequent low-value payments, and perform on-chain operations. However, these activities are unevenly distributed across networks, and registration scale does not necessarily correspond to reputation use or wallet activity. Current Web4 development is further hindered by missing security controls, documentation gaps, build failures, and limited diagnostic support, while delegated authorization, on-chain execution, and payment settlement remain critical Web4-specific issues. Projects address these problems through implementation changes, tests, documentation, and security hardening, but many reports still receive no technical response. Overall, Web4 is operational but not yet reliably engineered; stronger authorization controls, integration support, operational diagnostics, transaction handling, and payment-delivery guarantees are needed to build a dependable agent economy.

\bibliographystyle{ACM-Reference-Format}
\bibliography{references}
\end{document}